\documentclass[acus]{JAC2001}

\usepackage{graphicx,epsf}

\newcommand{\ie}{{\it i.e.}}
\newcommand{\vs}{{\it vs.}}
\begin{document}
\title{
\vspace{-.1in}\rightline{\rm\small IIT-HEP-01/3}
PROGRESS IN ABSORBER R\&D 2: WINDOWS\thanks{Presented at the 2001
Particle Accelerator Conference (PAC 2001), June 18--22, 2001, Chicago,
Illinois.} 
}

\author{D. M. Kaplan, E. L. Black, K. W. Cassel (Illinois Institute of 
Technology), S. Geer, \\
M. Popovic (Fermilab),
S. Ishimoto, K. Yoshimura (KEK), L. Bandura, M. A. Cummings,\\
A. Dyshkant, D. Kubik, D. Hedin 
(Northern Illinois 
Univ.), C. Darve (Northwestern Univ.),\\
Y. Kuno (Osaka
Univ.), D. Errede, M. Haney,
S. Majewski (Univ. of  Illinois
\\
at Urbana-Champaign), M. Reep, D. Summers (Univ. of Mississippi)}

\maketitle

\begin{abstract}
A program is underway to develop liquid-hydrogen energy absorbers for ionization cooling of muon-beam transverse emittance.  Minimization of multiple-scattering-induced  beam heating requires thin windows. The first window prototype has been destructively tested, validating the finite-element-analysis model and the design approach.
\end{abstract}

\section{INTRODUCTION}

High-energy stored muon beams may allow uniquely sensitive studies of  neutrino physics as well as compact lepton colliders for the study of the Higgs boson(s), supersymmetry, and the energy frontier~\cite{Status-Report,INSTR99}. An important technique for the creation of such beams is ionization cooling~\cite{Neuffer2,Fernow}, by which the transverse emittance of the initially-diffuse muon beam can be quickly reduced to dimensions commensurate with the acceptances of recirculating accelerators. Simulations show that enough transverse cooling can be achieved to build
a neutrino factory~\cite{FS2,Overarching-report}. We report here on recent progress in constructing prototype energy absorbers for muon-beam ionization cooling.

The Muon Collaboration (working with many additional physicists and engineers) has completed its second feasibility study of a neutrino factory based on a muon storage ring. The Feasibility Study II (FS2) report~\cite{FS2} describes a design that could be built for a well-defined cost, and that would deliver a flux of neutrinos for long-baseline neutrino-oscillation studies six times higher than that of the previous design iteration~\cite{FS1}.  Our next goals are to complete the designs for key muon-cooling components, build and test prototypes, and use them to carry out a first experimental demonstration of muon ionization cooling.

\section{ENERGY ABSORBERS FOR IONIZATION COOLING}

\begin{figure*}[t]
\vspace{-.3in}
\centerline{\includegraphics*[width=170mm]{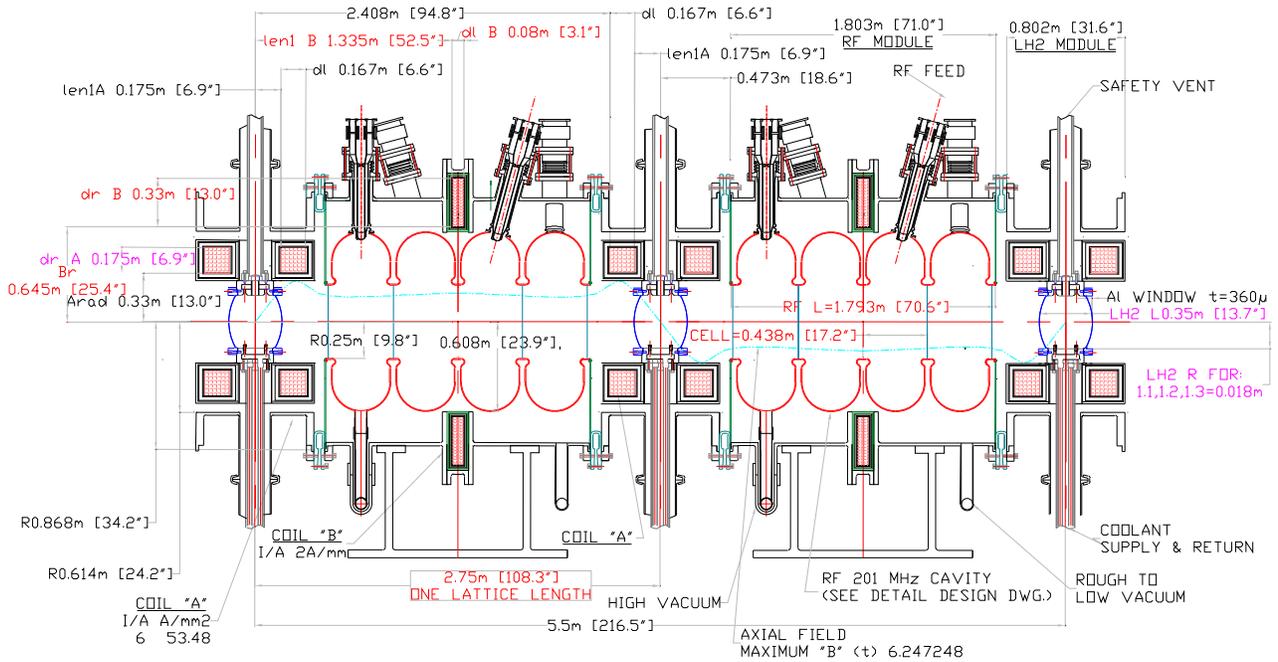}}
\vspace{-.3in}
\caption{Mechanical layout of  a portion of the ``SFOFO 1" ionization-cooling lattice from  the FS2 cooling channel, comprising two full cooling cells and part of a third. Shown are eight superconducting-solenoid coils, two 4-cell RF cavities, and three LH$_2$ absorbers.}
\label{fig:SFOFO}
\end{figure*}

Ionization cooling involves the damping of the muons' random transverse motions by ionization energy loss in an energy-absorbing medium; energy lost in the longitudinal direction is replaced via RF acceleration. From Eq.~\ref{eq1}~\cite{Neuffer2,Fernow},  maximizing the cooling rate $d\epsilon_n/ds$ requires minimizing the deleterious effects of Coulomb scattering in the absorbers by constructing them out of a material with a long radiation length ($L_R$) and embedding them in a focusing lattice (Fig.~\ref{fig:SFOFO}) at locations with as low  $\beta$ as possible:  
\begin{equation}
\frac{d\epsilon_n}{ds}\ =\
-\frac{1}{(v/c)^2} \frac{dE_{\mu}}{ds}\ \frac{\epsilon_n}{E_{\mu}}\ +
\ \frac{1}{(v/c)^3} \frac{\beta (0.014)^2}{2\ E_{\mu}m_{\mu}\ L_R}\,,
\label{eq1}
 \end{equation} 
where muon energy $E_\mu$  is in GeV,
$\epsilon_n$ is normalized emittance, and $s$ is path length.
In an optimized design the focusing is provided by superconducting solenoids and the absorber is liquid hydrogen (LH$_2$)~\cite{other-absorber}. 

The FS2 design includes absorbers of three types, as specified in Table~\ref{tab:FS2-abs}. 
The large transverse dimensions of the muon beam require large apertures and correspondingly wide absorbers, while the large energy spread of the beam demands frequent rebunching via RF cavities, favoring thin absorbers.  These two requirements lead to 
the slightly oblate shapes of the ``SFOFO" absorbers implied in Table~\ref{tab:FS2-abs} and shown in Fig.~\ref{fig:SFOFO}. 

\begin{table}
\begin{center}
\caption{LH$_2$ absorbers in Feasibility Study II.\label{tab:FS2-abs}}
\begin{tabular}{l|cccc}
\hline\hline
& Length & Radius & Number  & Power \\
{Absorber} &(cm) &(cm) & needed &(kW) \\
\hline
Minicooling & 175 & 30 & 2 & $\approx$5.5 \\
SFOFO 1 & 35 & 18 & 16 & $\approx $0.3 \\
SFOFO 2 & 21 & 11 & 36 & $\approx$0.1\\
\hline\hline
\end{tabular}
\end{center}
\end{table}

\subsection{Absorber Windows}

LH$_2$ containment requires a vessel with closed ends (windows) through which the muons must pass. A practical design choice for the windows is aluminum alloy.\footnote{Beryllium or a beryllium-containing alloy might be a superior choice, but beryllium has a questionable safety record in LH$_2$ applications.}  Simulations show that scattering in the absorber windows degrades muon-cooling performance. To keep this effect to a minimum, the FS2 design calls for window thicknesses as given in Table~\ref{tab:winthick}. (Note that in the long ``minicooling" absorbers, scattering is dominated by the hydrogen itself and thus the windows are not at issue).

\begin{table}
\begin{center}
\caption{Window thicknesses and operating pressures for the FS2 LH$_2$ absorbers.\label{tab:winthick}}
\begin{tabular}{l|cc}
\hline\hline
& Window thickness  & Max. operating\\
{Absorber}  & ($\mu$m) & pressure (atm) \\
\hline
Minicooling & -- & -- \\
SFOFO 1 & 360  & 1.2 \\
SFOFO 2 & 220 & 1.2 \\
\hline\hline
\end{tabular}
\end{center}
\end{table}

\begin{figure*}[t]
\centering
\epsfxsize=150mm\epsffile{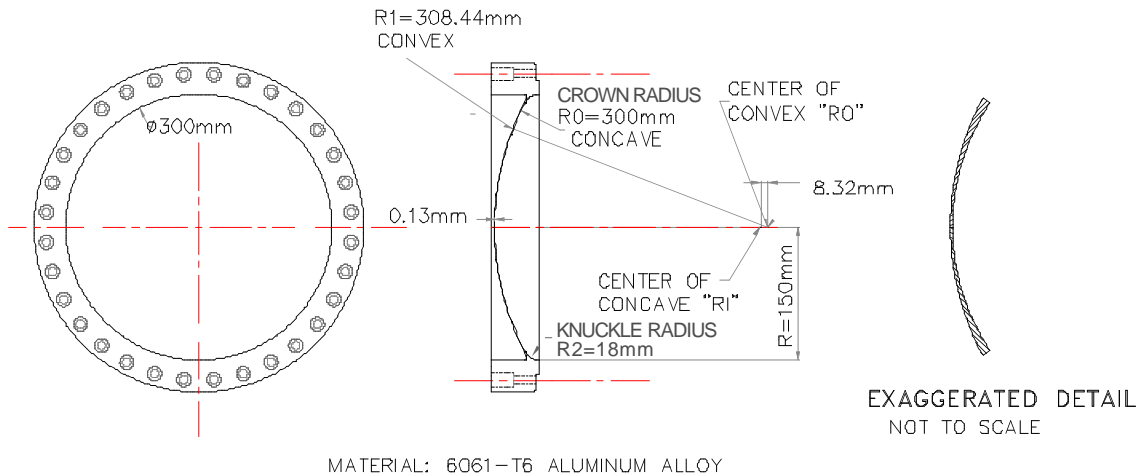}
\vspace{-2in}
\caption{Detail of tapered-torispherical window design.}
\label{fig:window}
\end{figure*}

Since the SFOFO absorbers are wider than they are long, hemispherical windows (which would be thinnest at a given pressure) are ruled out, and we are led to the ``torispherical" window shape. As specified by the American Society of Mechanical
Engineers (ASME)~\cite{ASME}, the torispherical head for pressure vessels is composed of a central portion having a radius of curvature (the ``crown radius") equal to the diameter of
the cylindrical portion of the vessel, joined to the cylindrical portion by a section of a toroidal surface with a radius of curvature 6\% of the crown radius (see Fig.~\ref{fig:window}). 

For an ASME-standard torispherical window, the required thickness is~\cite{ASME}
\begin{equation} 
t = \frac{0.885PD}{SE-0.1P}\,, \label{eq:winthick}
\end{equation} 
where $P$ is the pressure differential, $D$ the
length of the major axis (\ie\ the absorber diameter), 
$S$ the maximum allowable stress, and $E$ the
weld efficiency. 
For $S$, we follow ASME recommendations and
use the smaller of 1/4 of the ultimate strength $S_u$ or 2/3 of the yield strength $S_y$.\footnote{In
practice, for aluminum alloys, the ultimate strength provides the more stringent limit.} For 1.2-atm operation, and given the ASME specification for 6061-T6 aluminum alloy,\footnote{6061-T6 is the standard aluminum alloy for cryogenic applications, however more exotic high-strength alloys may also be suitable and are under investigation.} $S_u$ =289\,MPa, we obtain $t \ge 530\,\mu$m for the ``SFOFO 1" absorbers and $t \ge 330\,\mu$m for the ``SFOFO 2" absorbers, where the ``$>$" sign applies if the window is welded to its mounting flange ($E<1$).
However, to reach the smaller window thicknesses given in Table~\ref{tab:winthick}, we have devised a design in which each window is machined out of a single block of material, with an integral flange (with no welds, $E=1$), and the window thickness is tapered (based on finite-element analysis) to improve structural strength (Fig.~\ref{fig:window}).

\subsection{Window Prototype}

\begin{figure}[htb]
\centering
\includegraphics*[width=82mm]{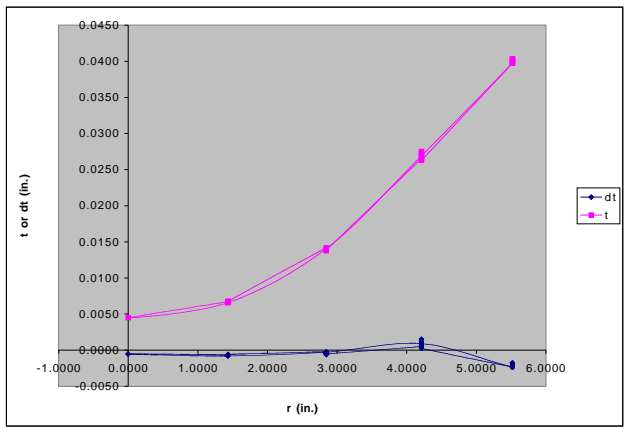}
\caption{Prototype-window measurements. Shown are thickness $t$ and thickness error $dt$ \vs\ radius, all in inches.}
\label{fig:meas}
\end{figure}

We have built and tested a first prototype window of the above design. To test the limits of the proposed manufacturing technique, we specified a central thickness of only 130\,$\mu$m, with a radius of 15\,cm. The window was built on a CNC milling machine and CNC lathe at the Univ.\ of Mississippi. After one face was machined, a custom-built backing jig was used to support it while the other face was cut.
The window was then measured using a precision coordinate-measuring machine at Fermilab and with micrometers and found to be within 5\% of the nominal thickness profile (Fig.~\ref{fig:meas}), validating the design approach and manufacturing procedure.

\subsection{Pressure Tests}

To be certified as safe for  liquid-hydrogen containment at Fermilab, the vessel must undergo a stringent safety review. The requirements include destructive testing of five windows of a given design before a sixth window may be put into service. The first prototype was pressure-tested in a setup in which the window, with 22 strain gages affixed at strategically-chosen points on its surface, was mounted to a back plate, the volume thereby enclosed being filled with water. The water was then pressurized to varying degrees with air and the resulting strains read out to a PC via a scanning DVM and 22 ADC channels. Additional measurements included the pressure, the volume of water contained, and precision photogrammetric measurements\footnote{Photogrammetry is attractive in that it permits non-contact monitoring of strain. We are not aware of its prior use for such a purpose. Contact measurements are undesirable since the gluing on of strain gages is  labor intensive and (especially with such a thin foil) may bias the  measurement.}
of the shape of the window surface. Detailed
comparison of these measurements with the predictions of a finite-element-analysis model will allow the design and fabrication procedures to be certified for future windows of various  sizes and thicknesses. 

Summarized briefly, at the present stage of analysis, the agreement between the photogrammetric measurements and the strain-gage data is good, with typical discrepancies below 10\%.  The window-failure mode was somewhat surprising: while the onset of inelastic deformation was predicted to occur at 29\,psig, a pinhole leak appeared at 31\,psig, probably due to a defect. Massive rupture ensued at 44\,psig. More detailed results will appear in a forthcoming publication. 

\section{Acknowledgements}

We thank R. Riley for measuring the prototype window after manufacture, P. Stone for mechanical support, and J. Greenwood for carrying out the photogrammetric survey. This work was
supported in part by the U.S. Dept.\ of Energy, the National Science
Foundation, the Illinois Board of Higher Education, and the Illinois Dept.\ of Commerce and Community Affairs.


\end{document}